\documentclass[twocolumn,aps,superscriptaddress,showpacs,nofootinbib,floatfix]{revtex4-1}
\usepackage{graphicx} % Required for inserting images
\usepackage{epsfig,bm}
\usepackage{graphicx,amssymb}
\usepackage{amsmath}
\usepackage{epstopdf}
\usepackage[eps]{pstricks}
\usepackage{slashed}

\usepackage[normalem]{ulem}

\renewcommand{\sout}{\bgroup \color{red} \ULdepth=-.5ex \ULset}

\begin{document}

\title{Is $K_{1}/K^{*}$ enhancement in heavy ion collisions a signature for chiral symmetry restoration?}

\author{Haesom Sung}
\email{ioussom@yonsei.ac.kr}
\affiliation{Department of Physics and Institute of Physics and Applied Physics, Yonsei University, Seoul 03722, Korea}
\affiliation{Cyclotron Institute, Texas $A\&M$ University, College Station, TX 77843, USA}

\author{Sungtae Cho}\email{sungtae.cho@kangwon.ac.kr}
\affiliation{Division of Science Education, Kangwon National University, Chuncheon 24341, Korea}

\author{Che Ming Ko}\email{ko@comp.tamu.edu}
\affiliation{Cyclotron Institute, Texas $A\&M$ University, College Station, TX 77843, USA}
\affiliation{Department of Physics and Astronomy, Texas $A\&M$ University, College Station, TX 77843, USA}

\author{Su Houng Lee}\email{suhoung@yonsei.ac.kr}
\affiliation{Department of Physics and Institute of Physics and Applied Physics, Yonsei University, Seoul 03722, Korea}

\author{Sanghoon Lim}\email{shlim@pusan.ac.kr}
\affiliation{Department of Physics, Pusan National University, Pusan, Republic of Korea}

\begin{abstract}
We extend the recent study of $K_{1}/K^{*}$ enhancement as a signature of chiral symmetry restoration in heavy ion collisions at the Large Hadron Collider (LHC) via the kinetic approach to include the effects due to non-unity hadron fugacities during the evolution of produced hadronic matter and the temperature-dependent $K_1$ mass.  Although including non-unity pion and kaon fugacities reduces slightly the $K_1/K^*$ enhancement found in previous study due to chiral symmetry restoration, adding temperature-dependent $K_1$ mass leads to a substantial further reduction of the $K_1/K^*$ enhancement.  However, the final $K_1/K^*$ ratio in peripheral collisions still shows a factor of 2.4 enhancement compared to the case without chiral symmetry restoration, confirming its use as a good signature for chiral symmetry restoration in the hot dense matter produced in relativistic heavy ion collisions.
\end{abstract}

\maketitle

\section{Introduction}

According to lattice QCD calculations, the quark-gluon plasma (QGP) to hadronic matter (HM) transition at vanishing baryon chemical potential is a smooth crossover with a critical temperature $T_C$ at about 156 MeV~\cite{HotQCD:2018pds}. This temperature coincides with the chemical freeze-out temperature in the statistical model for particle production in relativistic heavy ion collisions at energies available from the Relativistic Heavy Ion Collider (RHIC) and the LHC~\cite{Andronic:2005yp,Andronic:2012dm, Stachel:2013zma}.   Since the chiral symmetry is restored above this temperature, masses of chiral partners are expected to become degenerate near $T_C$ as indicated in studies based on the QCD sum rules for the axial vector meson $K_{1}(1270)$ and vector meson $K^*(890)$ masses~\cite{Lee:2023ofg} as well as the lattice QCD~\cite{Skullerud:2022yjr} and the functional renormalization group~\cite{Jung:2016yxl} calculations for the axial vector meson $a_1(1260)$ and vector meson $\rho(770)$ masses. Because of the shorter lifetimes of $K_{1}(1270)$ and $K^*(890)$, which have vacuum decay widths of 90 MeV and 47 MeV, respectively, than that of the hadronic stage of relativistic heavy ion collisions, their yield ratio $K_1/K^*$ in these collisions is expected to depend on the degree of chiral symmetry restoration in the produced matter.  A recent study by some of the present authors~\cite{Sung:2021myr} has indeed found this effect in Pb+Pb collisions at $\sqrt{s_{NN}}=5.02$ TeV.  Using the $K_1$ number at $T_C$ obtained from the statistical hadronization model by taking the masses of $K_1$ and $K^*$ to be $m_{K_1}=m_{K^*}=890$ MeV according to a QCD sum rule calculation~\cite{Kim:2020zae} and assuming that the $K_1$ mass immediately changes to its vacuum mass in the produced hadronic matter, they have studied the effect of hadronic scatterings on the yield ratio $K_1/K^*$ via a kinetic approach.  Based on a schematic hydrodynamic model for the evolution of produced hot dense matter using the lattice equation of state for the QGP and the resonance hadron gas model for the HM~\cite{Song:2010fk}, the time evolution of $K_1$ and $K^*$ numbers are studied by taking into account the reactions $K_1\pi\leftrightarrow K\pi$, $K_1\pi\leftrightarrow K^*\rho$, $K_1\rho\leftrightarrow K^*\pi$, $K_1\rho\leftrightarrow  K\rho$, $K_1\leftrightarrow K^*\pi$ and $K_1\leftrightarrow K\rho$ that involve the $K_1$ meson as well as the reactions $K^*\pi\leftrightarrow K\rho$, $K^*\rho\leftrightarrow K\pi$ and $K^*\leftrightarrow K\pi$ that involve the $K^*$ meson.  Their results show that the ratio $K_1/K^*$ is increased by a factor of 3 in mid-central collisions (40-50\% centrality) and by a factor of 6 in peripheral collisions (70-80\% centrality) compared to that without including the effect of chiral symmetry restoration, although it is not affected much in central collisions (0-5\% centrality).

The study in Ref.~\cite{Sung:2021myr} has, however, neglected two important effects, namely, 1) the constancy of effective pion, kaon and nucleon numbers during the hadronic evolution after including those from resonance decays, which is supported by the success of the statistical hadonization model that these effective numbers are fixed at $T_C$ when the chemical freeze out takes place, and 2) the temperature-dependent $K_1$ mass in the hadronic matter ~\cite{Lee:2023ofg}. As shown in a study based on a multi-phase transport (AMPT) model~\cite{Xu:2017akx}, constant effective pion, kaon and nucleon numbers is accompanied by a constant entropy per particle during the hadronic evolution, indicating non-unity pion, kaon and nucleon fugacities if the hadronic matter is modeled by a thermally equilibrated fireball that cools as it expands. In the present study, we extend the study of Ref.~\cite{Sung:2021myr} to include this effect and also the temperature-dependent $K_1$ mass given in Ref.~\cite{Lee:2023ofg} by using the temperature-dependent quark condensate from Ref.~\cite{Weise:2012yv}.  Including these two effects in the kinetic equations allow us to study more realistically the $K_1/K^*$ ratio in relativistic heavy ion collisions.  Although results from present study show a smaller $K_1/K^*$ ratio than in Ref.~\cite{Sung:2021myr}, they do not change the conclusion that an enhanced $K_1/K^*$ ratio than that predicted by the statistical hadronization model can serve as a good signature for the chiral symmetry restoration in the hot dense matter produced in relativistic heavy ion collisions.

The present paper is organized as follows. We first review in Sec.~\ref{mass} the temperature dependence of $K_1$ mass in a hadronic matter at finite temperature and then use it in Sec.~\ref{crosssection} to calculate the cross sections for $K_1$ and $K^*$ reactions with pion and rho meson as well as their thermal averages.  In Sec.~\ref{fugacities}, we determine the temperature dependence of the pion, kaon, and nucleon fugacities by requiring the effective pion, kaon and nucleon numbers, which included those from resonance decays, as well as the entropy per particle to remain unchanged during the hadronic evolution. The kinetic equations for the time evolution of the $K_{1}$ and $K^*$ numbers are then given in Sec.~\ref{kinetic}, with the results on the yield ratio $K_1/K^*$ in Pb+Pb collisions presented in Sec.~\ref{results}.  Finally, a brief summary is given in Sec.~\ref{summary}.   

\section{Temperature-dependent $K_{1}$ meson mass}\label{mass}

\begin{figure}[h]
\centering
\includegraphics[width=0.9\linewidth]{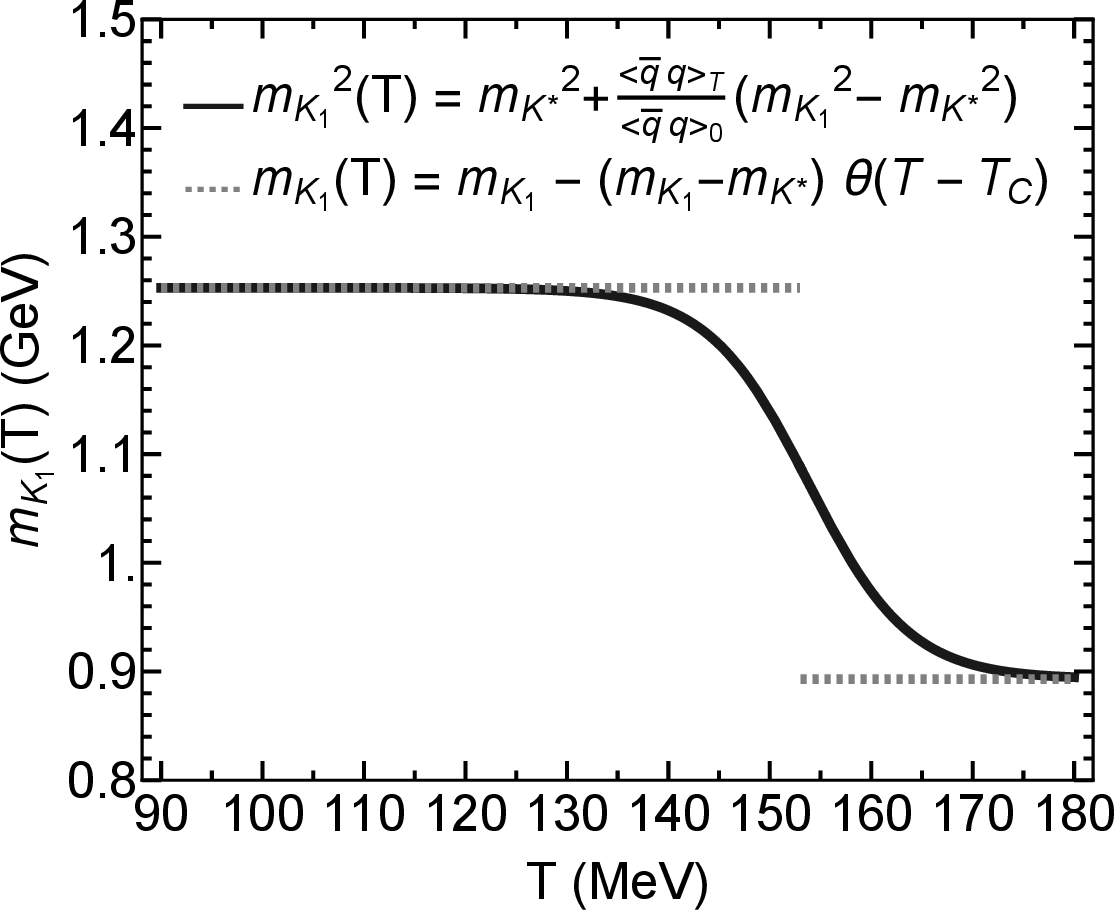}
\caption{Temperature dependence of $K_{1}$ mass.  Solid line is from the QCD sum rule calculations of  Ref.~\cite{Lee:2023ofg}, while dotted line is the one assumed in Ref.~\cite{Sung:2021myr} with $T_C=156$ MeV.}
\label{K1T}
\end{figure}

According to the QCD sum rule study of Ref.~\cite{Lee:2023ofg}, the mass difference between $K_{1}$ and $K^*$ mesons in a hot hadronic matter depends on the quark condensate $\langle \bar{q}q\rangle_{T}$ as 
\begin{equation}\label{eq:mass_K1}
    m_{K_{1}}^{2}(T) = m_{K^{*}}^{2}+\frac{\langle \bar{q}q\rangle_{T}}{\langle \bar{q}q\rangle_{0}}(m_{K_{1}}^{2}-m_{K^{*}}^{2}),
\end{equation}
where $\langle\bar q q\rangle_0$ is the quark condensate in the vacuum. Neglecting the small change of $K^*$ mass with temperature~\cite{Kim:2020zae} and using  $m_{K_{1}}$=1.25 GeV, $m_{K^{*}}$=0.892 GeV, and the temperature-dependent quark condensate from Ref.~\cite{Weise:2012yv}, the temperature dependence of $K_1$ mass is shown in Fig.~\ref{K1T}. It is seen that the $K_1$ mass at $T_C$ is about 1.1 GeV, instead of the $K^*$ free-space mass of 0.892 GeV assumed in Ref.~\cite{Sung:2021myr}, and then gradually increases to its free-space value of 1.25 GeV.  

\section{$K_{1}$ and $K^{*}$ reaction cross sections}\label{crosssection}

\begin{figure}[h]
\centering
\includegraphics[width=0.9\linewidth]{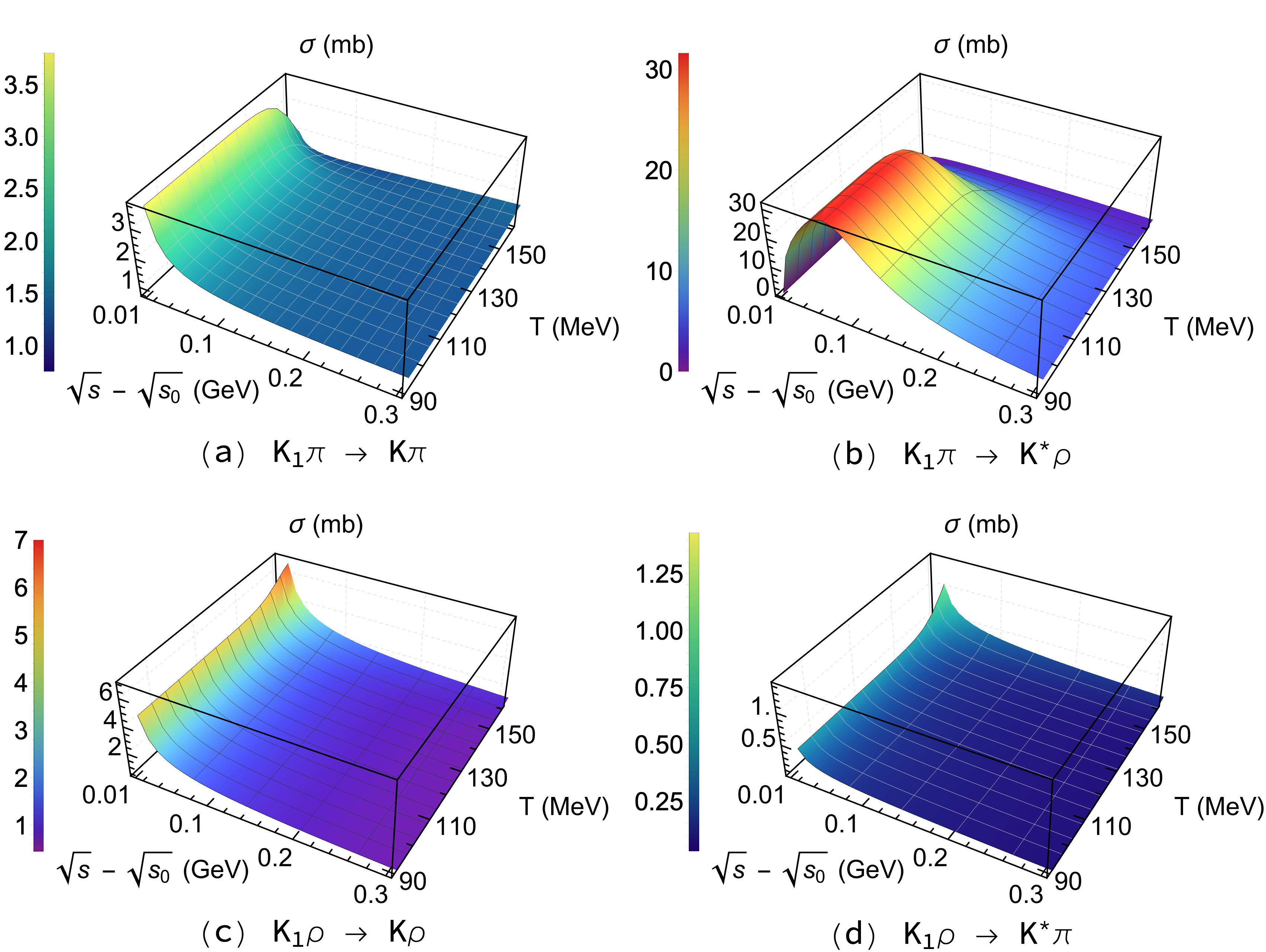}
\caption{Isospin averaged cross sections for $K_{1}\pi \to K \pi$ (panel (a)), $K_{1}\pi\to K^{*}\rho$ (panel (b)), $ K_{1}\rho \to K\rho$ (panel (c)), and $K_{1}\rho\to K^{*}\pi$ (panel (d)) as functions of center-of-mass energy $\sqrt{s}$ with $\sqrt{s_0}$ denoting the threshold of a reaction and temperature.}
\label{cross}
\end{figure}

In this Section, we review the $K_1$ and $K^*$ reaction cross sections with pion and rho meson, whose abundance dominate in the hadronic matter.  These reactions include $K_{1}+ \pi \to K + \pi$, $K_{1}+ \pi \to K^{*}+ \rho$, $K_{1}+\rho \to K + \rho$, and $K_{1}+ \rho \to K^{*}+ \pi$ for the $K_1$ meson, and their cross sections have been calculated in Ref.~\cite{Sung:2021myr} using the massive Yang-Mills approach with a Lagrangian involving spin-0 and spin-1 mesons~\cite{Meissner:1987ge}. Shown in Fig.~\ref{cross} are the center-of-mass energy $\sqrt{s}$ and temperature dependence of their isospin averaged cross sections.  The most important channel for $K_1$ annihilation is the endothermic reaction $K_{1}+ \pi \to K^* + \rho$, except near its threshold where other reactions dominate because of their exothermic nature. In calculating the pion-exchange $t$-channel diagram in the reaction $K_{1}+ \pi \to K^* + \rho$, the pion can be on shell at certain reaction energy.  In this case, the reaction $K_{1}+ \pi \to K^* + \rho$ is the same as the two-step process of $K_1\to K^*+\pi$ followed by $\pi+\pi\to\rho$.  Since the process $K_1\to K^*+\pi$ is explicitly included in the kinetic equations used in our study, we therefore exclude the contribution of on-shell pion to the pion-exchange $t$-channel diagram of the reaction $K_{1}+ \pi \to K^* + \rho$ as in Ref.~\cite{Sung:2021myr}.   

\begin{figure}[h]
\centering
\includegraphics[width=0.9\linewidth]{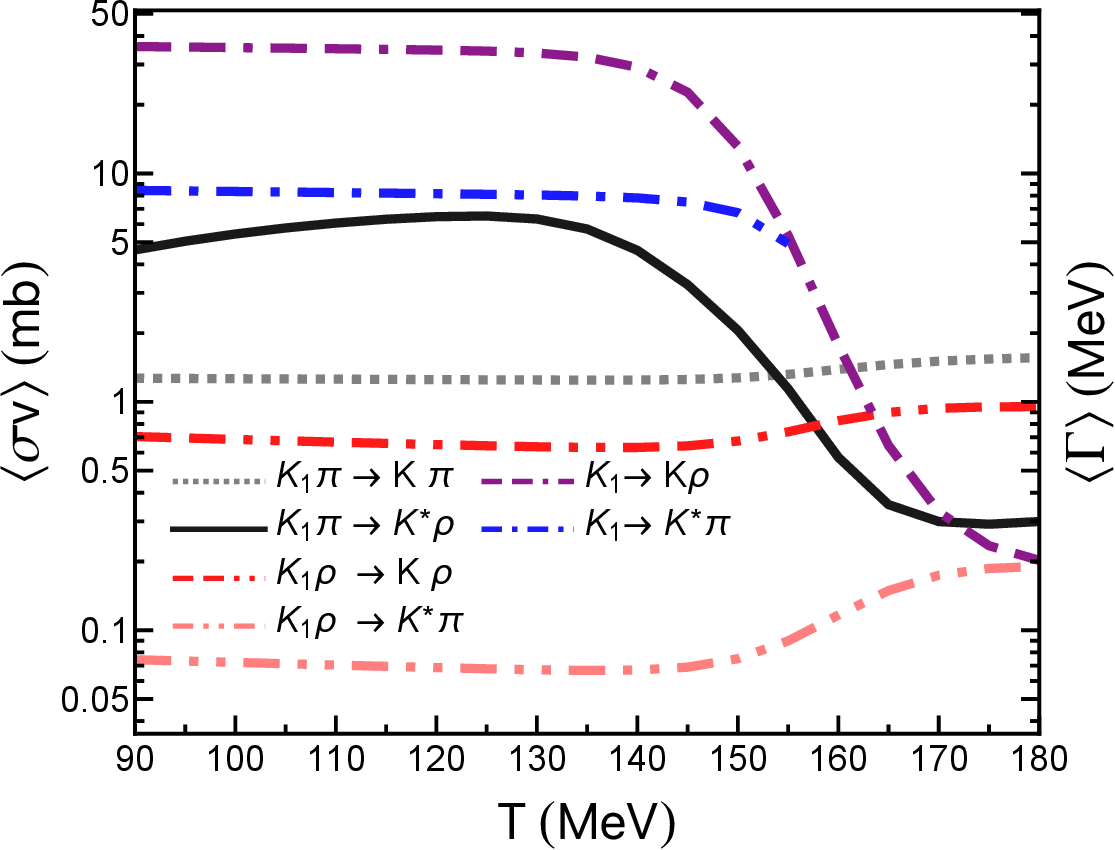}
\caption{Temperature dependence of thermal averaged cross sections $\langle\sigma v\rangle$ for the reactions $K_{1}\pi \to K \pi$ (dotted line), $K_{1}\pi\to K^{*}\rho$ (solid line), $ K_{1}\rho \to K\rho$ (dash-dash-dot-dotted line), and $K_{1}\rho\to K^{*}\pi$ (dash-dot-dotted line), and thermal averaged decay widths $\langle\Gamma\rangle$ of $K_1\to K\rho$ (dash-dash-dotted line) and $K_1\to K^*\pi$ (dash-dotted line). }
\label{K1}
\end{figure}

The above reactions enter the kinetic equations, which are given in Sec.~\ref{kinetic}, through their thermal average over the momentum distributions of the particles in the initial state, i.e., 
\begin{eqnarray}\label{eq:AvgCrs}
\left<\sigma_{ab\rightarrow cd}v_{ab}\right> &=& \frac{\int d^{3}{\bf p}_{a}d^{3}{\bf p}_{b}f_{a}({\bf p}_{a})f_{b}({\bf p}_{b})\sigma_{ab\rightarrow cd}v_{ab}}{\int d^{3}{\bf p}_{a}d^{3}{\bf p}_{b}f_{a}({\bf p}_{a})f_{b}({\bf p}_{b})}.
\end{eqnarray}
In the above, $f_i({\bf p}_i)$ is the Boltzman momentum distribution of particle species $i=a,b$, i.e., $f_i({\bf p}_i)=e^{-\sqrt{{\bf p}_i^{2}+m_i^{2}}/T}$ with $m_i$ being the particle mass, which we take as their vacuum masses for pion, kaon, rho meson, and $K^*$ and the temperature-dependent mass for $K_1$. The $v_{ab}$ in the above equation is the relative velocity between the two initial particles $a$ and $b$. The temperature-dependent thermal averaged cross sections for $K_{1}$ annihilation by pion and rho meson are shown in Fig.~\ref{K1}, where it is seen that $\langle\sigma_{K_1\pi\to K^*\rho}\rangle$ dominates over other thermal averaged cross sections at the temperature range of interest for the present study.   Also shown in Fig.~\ref{K1} are the thermal averaged decay widths of $K_1$ meson to $K\rho$ and $K^*\pi$, which are computed according to $\left<\Gamma_{K_{1}}\right>=\Gamma_{K_{1}}(m_{K_1})K_1(m_{K_1}/T)/K_2(m_{K_1}/T)$ with $\Gamma_{K_{1}}(m_{K_1})$ evaluated with the inclusion of the $\rho$ mass distribution in the final state, where $K_1(x)$ and $K_2(x)$ are modified Bessel functions of first and second kind, respectively, to take into account its temperature-dependent mass and the effect of time dilation. The $\langle\Gamma_{K_1\to K\rho}\rangle$ is seen to have a larger value than $\langle\Gamma_{K_1\to K^*\pi}\rangle$.

For the $K^{*}$ annihilation processes, they include the reactions $K^*\pi\to K\rho$ and $K^*\rho\to K\pi$ and the decay process $K^*\to K\pi$.  Their values and thermal averages have been calculated in Ref.~\cite{Cho:2015qca} by using the free-space $K^*$ mass, which we will use since we also neglect the small temperature dependence of the $K^*$ mass in the present study. 

\section{Fugacities of pion, kaon and nucleon}\label{fugacities}

\begin{figure}[h]
\centering
\includegraphics[width=0.8\linewidth]{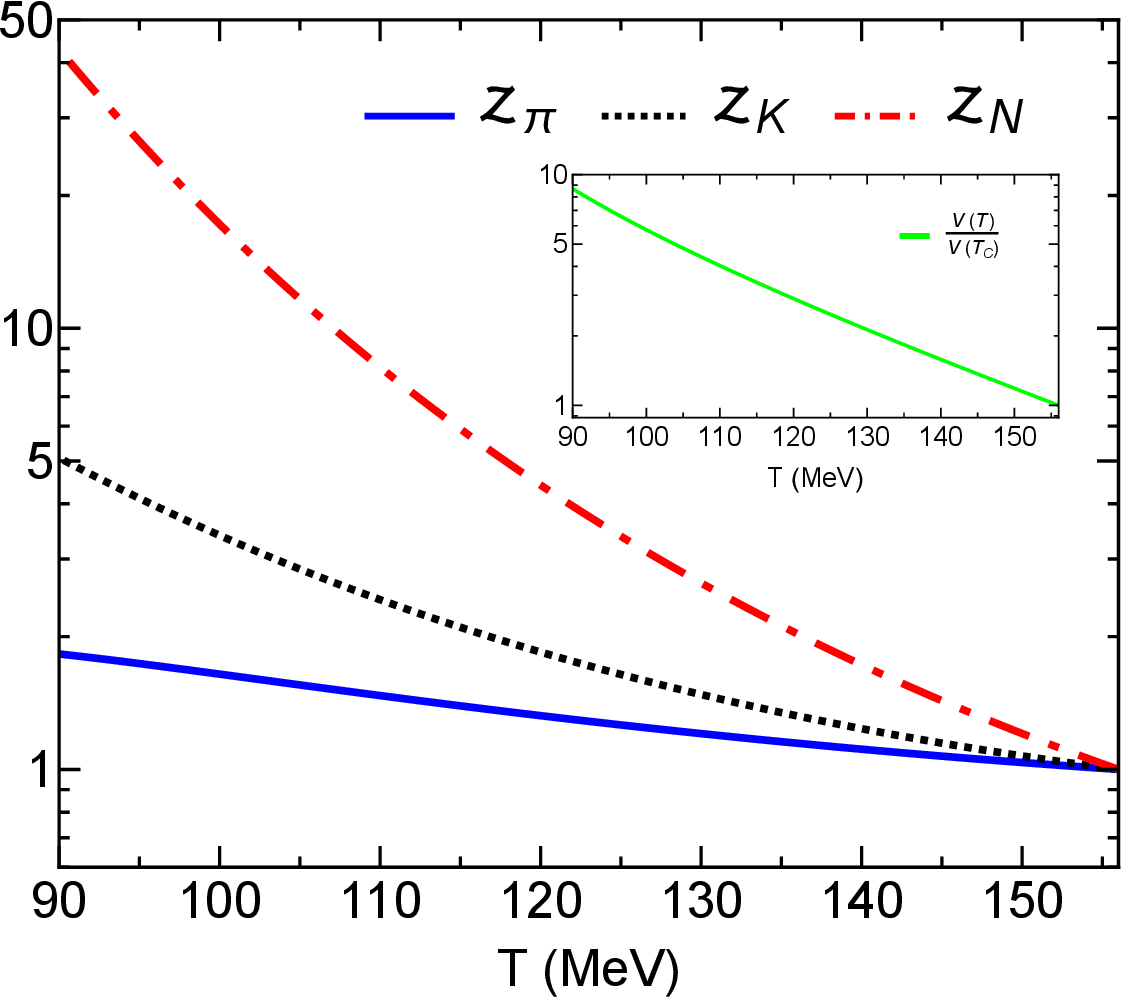}
\caption{Temperature dependence of pion (solid line), kaon (dashed line) and nucleon (dash-dotted line) fugacities as well as the volume ratio of hadronic matter (solid line in the inset).}
\label{fugacity}
\end{figure}

According to the statistical model for particle production in relativistic heavy ion collisions, particle yields including contributions from resonances decays, i.e., their effective numbers, are determined at the chemical freeze-out temperature, which coincides with the QGP to HM phase transition temperature~\cite{Andronic:2005yp,Andronic:2012dm, Stachel:2013zma}.  To maintain the effective pion, kaon and nucleon numbers, which are relevant in the present study, during the expansion and cooling of the hadronic matter, it is necessary for them to acquire non-unity fugacity, as shown in Ref.~\cite{Xu:2017akx}.   In this case, the pion, kaon and nucleon momentum distributions in the Boltzmann approximation need to be multiplied by their fugacity $z_i$, i.e., $z_if_i({\bf p})$.  In terms of the thermally equilibrated density $n_i^T=\frac{g_i}{(2\pi)^3}\int d^3{\bf p} f_i({\bf p})$ of particle species $i$, where $g_i$ is its spin and isospin degeneracies, the effective pion, kaon and nucleon densities in a hadronic matter of temperature $T$ is then given by the sum of the densities of free pions, kaons, and nucleons as well as those from resonance decays, i.e.,  
\begin{eqnarray}
&&n^{\rm eff}_{\pi}(T) = z_\pi n_{\pi}^T + z_{\pi}^{2}n_{\rho}^T+z_{\pi}z_{K}n_{K^{*}}^T+z_\pi^2 z_K n_{K_1}^T\notag\\
&&~~~~~~~~~~~~+z_\pi z_Nn_\Delta^T+\cdots,\\
&&n^{\rm eff}_{K}(T) = z_K n_K^T + z_{\pi}z_Kn_{K^*}^T+z_{\pi}^2z_{K}n_{K_1}^T+z_K^2 n_\phi^T\notag\\
&&~~~~~~~~~~~~+\cdots,\\
&&n^{\rm eff}_{N}(T) = z_N n_N^T+z_\pi z_Nn_\Delta^T+\cdots.
\end{eqnarray}
In the above, $\cdots$ denotes the contribution from strong decays of other resonances, which we include all particles of masses up to 2 GeV in the particle data book.  In obtaining the above equations, we have also used the relations $z_\rho=z_\pi^2$, $z_{K^*}=z_\pi z_K$, $z_{K_1}=z_\pi^2z_K$, $z_\Delta=z_\pi z_N$, etc. from the assumption that all particles are in thermal and chemical equilibrium.  In terms of the pion, kaon and nucleon fugacities, the entropy and particle densities of a hadronic matter at temperature $T$ are then given by
\begin{eqnarray}
&&s(T)=-\sum_i g_i\int\frac{d^3{\bf p}}{(2\pi)^3} (z_if_i)\ln (z_if_i),\\
&&n(T)=\sum_i z_in_i^T,
\end{eqnarray}
where the summation $i$ again includes all particles of masses up to 2 GeV. As shown in Eq.(6), the relativistic Boltzmann distribution is used to evaluate the entropy density as in the calculation of the thermal averaged cross sections and decay widths given by Eq.(2), the effective pion, kaon and nucleon densities in Eqs.(3)-(5) as well as in the total particle density in Eq.(7).

Starting with an initial temperature $T_C$ and volume $V_C$ at hadronization of the QGP produced in relativistic heavy ion collisions, when all particles have unity forgacities according to the statistical model for particle production, the volume $V(T)$ of the hadronic matter and the pion, kaon and nucleon furgacities $z_\pi$, $z_K$ and $z_N$ at a later time when the temperature drops to $T$ can be obtained from the constancy of entropy per particle and the effective pion, kaon and nucleon numbers by solving the four equations, $n_{\pi, K, N}^{\rm eff}V(T)=n_{\pi, K, N}^{\rm eff}(T_C)V(T_C)$ and $s(T)/n(T)=s(T_C)/n(T_C)$.   In Fig.~\ref{fugacity}, we show the temperature dependence of $z_\pi$, $z_K$, $z_N$ and $V(T)/V(T_C)$.  It is seen that their values all increase with decreasing temperature of the hadronic matter, with $z_N$ increasing faster than $z_K$ and $z_K$ increasing faster than $z_\pi$.  We note that the constant entropy per particle in the hadronic matter has a value of 6.1.  

\section{Kinetic equations for $K_{1}$, $K^{*}$ and $K$}\label{kinetic}

Neglecting the creation and annihilation of strange hadrons, such as the reaction $\pi\pi \leftrightarrow K\bar{K}$, which has little effect on the results in the present study, then $N_{0} = N_{K_{1}}+N_{K^{*}}+N_{K}$ is a constant during the hadronic evolution. In this case, the kinetic equation for the time evolution of $K_{1}$ number can be written as
\begin{eqnarray}\label{Kin_K1}
\frac{dN_{K_1}}{dt}=\gamma_{K_1,K_1}N_{K_1}+\gamma_{K_1,K^*}N_{K^*}+\gamma_{K_1,K}N_K,
\end{eqnarray}
where
\begin{eqnarray}
    &&\gamma_{K_1,K_1}=-(\langle\sigma_{K_1\pi\to K\pi}\rangle+\langle\sigma_{K_1\pi\to K^*\rho}v\rangle)z_\pi n_\pi^T\notag\\
    &&\qquad\qquad -(\langle\sigma_{K_1\rho\to K^*\pi}v\rangle+\langle\sigma_{K_1\rho\to K\rho}v\rangle)z_\pi^2 n_\rho^T\notag\\
    &&\qquad\qquad -\langle\Gamma_{K_1\to K^*\pi}\rangle-\langle\Gamma_{K_1\to K\rho}\rangle,\\
    &&\gamma_{K_1,K^*}=\langle\sigma_{K^*\rho\to K_1\pi}v\rangle z_\pi^2n_\rho^T\notag\\
    &&\qquad\qquad+(\langle\sigma_{K^*\pi\to K_1\rho}v\rangle+\langle\sigma_{K^*\pi\to K_1}v\rangle) z_\pi n_\pi^T\\
    &&\gamma_{K_1,K}=\langle\sigma_{K\pi\to K_1\pi}v\rangle z_\pi n_\pi^T,\notag\\
    &&\qquad\qquad+(\langle\sigma_{K\rho\to K_1\rho}v\rangle+\langle\sigma_{K\rho\to K_1}v\rangle) z_\pi^2 n_\rho^T,
\end{eqnarray}
with $n_\pi^T$, $n_\rho^T$, $n_K^T$, $n_{K^*}^T$ and $n_{K_1}^T$ being, respectively, the thermally equilibrated densities of $\pi$, $\rho$, $K$, $K^*$ and $K_1$ mesons. For the thermal averaged cross sections in Eqs.(10) and (11), which describe the regeneration of $K_1$ meson, they are related to the thermal averaged cross sections and decay widths in Eq.(9), which describe the annihilation of $K_1$ meson, by $\langle\sigma_{K^*\rho\to K_1\pi}v\rangle=\langle\sigma_{ K_1\pi\to K^*\rho}v\rangle\frac{n_{K_1}^Tn_\pi^T}{n_{K^*}^Tn_\rho^T}$, $\langle\sigma_{K^*\pi\to K_1\rho}v\rangle=\langle\sigma_{K_1\rho\to K^*\pi}v\rangle\frac{z_\pi^2 n_{K_1}^Tn_\rho^T}{n_{K^*}^Tn_\pi^T}$, $\langle\sigma_{K^*\pi\to K_1}v\rangle=\langle\Gamma_{K_1\to K^*\pi}\rangle\frac{n_{K_1}^T}{n_{K^*}^Tn_\pi^T}$, $\langle\sigma_{K\pi\to K_1\pi}v\rangle=\langle\sigma_{K_1\pi\to K\pi}v\rangle\frac{z_\pi^2n_{K_1}^T}{n_K^T}$, $\langle\sigma_{K\rho\to K_1\rho}v\rangle=\langle\sigma_{K_1\rho\to K\rho}v\rangle\frac{z_\pi^2n_{K_1}^T}{n_K^T}$, and $\langle\sigma_{K\rho\to K_1}v\rangle=\langle\Gamma_{K_1\to K\rho}\rangle\frac{n_{K_1}^T}{n_K^Tn_\rho^T}$.

Similarly, the kinetic equation for the time evolution of $K^*$ number is given by
\begin{eqnarray}\label{Kin_Ks}
\frac{dN_{K^*}}{dt}=\gamma_{K^*,K_1}N_{K_1}+\gamma_{K^*,K^*}N_{K^*}+\gamma_{K^*,K}N_K,
\end{eqnarray}
where
\begin{eqnarray}
    &&\gamma_{K^*,K_1}=\langle\sigma_{K_1\pi\to K^*\rho}v\rangle z_\pi n_\pi^T+\langle\sigma_{K_1\rho\to K^*\pi}v\rangle z_\pi^2n_\rho^T\notag\\
    &&\qquad\qquad +\langle\Gamma_{K_1\to K^*\pi}\rangle,\\ 
    &&\gamma_{K^*,K^*}=-(\langle\sigma_{K^*\pi\to K\rho}v\rangle+\langle\sigma_{K^*\pi\to K_1\rho}v\rangle\notag\\
    &&\qquad\qquad+\langle\sigma_{K^*\pi\to K_1}v\rangle)z_\pi n_\pi^T\notag\\
     &&\qquad\qquad-(\langle\sigma_{K^*\rho\to K\pi}v\rangle+\langle\sigma_{K^*\rho\to K_1\pi}v\rangle) 
     z_\pi^2n_\rho^T\notag\\
    &&\qquad\qquad -\langle\Gamma_{K^*\to K\pi}\rangle,\\
    &&\gamma_{K^*,K}=(\langle\sigma_{K\pi\to K^*\rho}v\rangle+\langle\sigma_{K\pi\to\ K^*}v\rangle)z_\pi n_\pi^T\notag\\
    &&\qquad\qquad+\langle\sigma_{K\rho\to K^*\pi}v\rangle z_\pi^2 n_\rho^T.
\end{eqnarray}
As in the case for the $K_1$ meson, the thermal averaged cross sections $\langle\sigma_{K\pi\to K^*\rho}v\rangle$, $\langle\sigma_{K\rho\to K^*\pi}v\rangle$, and $\langle\sigma_{K\pi\to\ K^*}v\rangle$ in Eq.(15) are related to the thermal averaged cross sections $\langle\sigma_{K^*\rho\to K\pi}v\rangle$ and $\langle\sigma_{K^*\pi\to K\rho}v\rangle$ and the thermal averaged width $\langle\Gamma_{K^*\to K\pi}\rangle$ in Eq.(14), which we take from Ref.~\cite{Cho:2015qca}, by $\langle\sigma_{K\pi\to K^*\rho}v\rangle=\langle\sigma_{K^*\rho\to K\pi}v\rangle\frac{z_\pi^2n_{K^*}^Tn_\rho^T}{n_K^Tn_\pi^T}$, $\langle\sigma_{K\rho\to K^*\pi}v\rangle=\langle\sigma_{K^*\pi\to K\rho}v\rangle\frac{n_{K^*}^Tn_\pi^T}{n_K^Tn_\rho^T}$, $\langle\sigma_{K\pi\to\ K^*}v\rangle=\langle\Gamma_{K^*\to K\pi}\rangle\frac{n_{K^*}^T}{n_K^Tn_\pi^T}$. 

\section{results}\label{results}

\begin{figure}[h]
\centering
\includegraphics[width=0.9\linewidth]{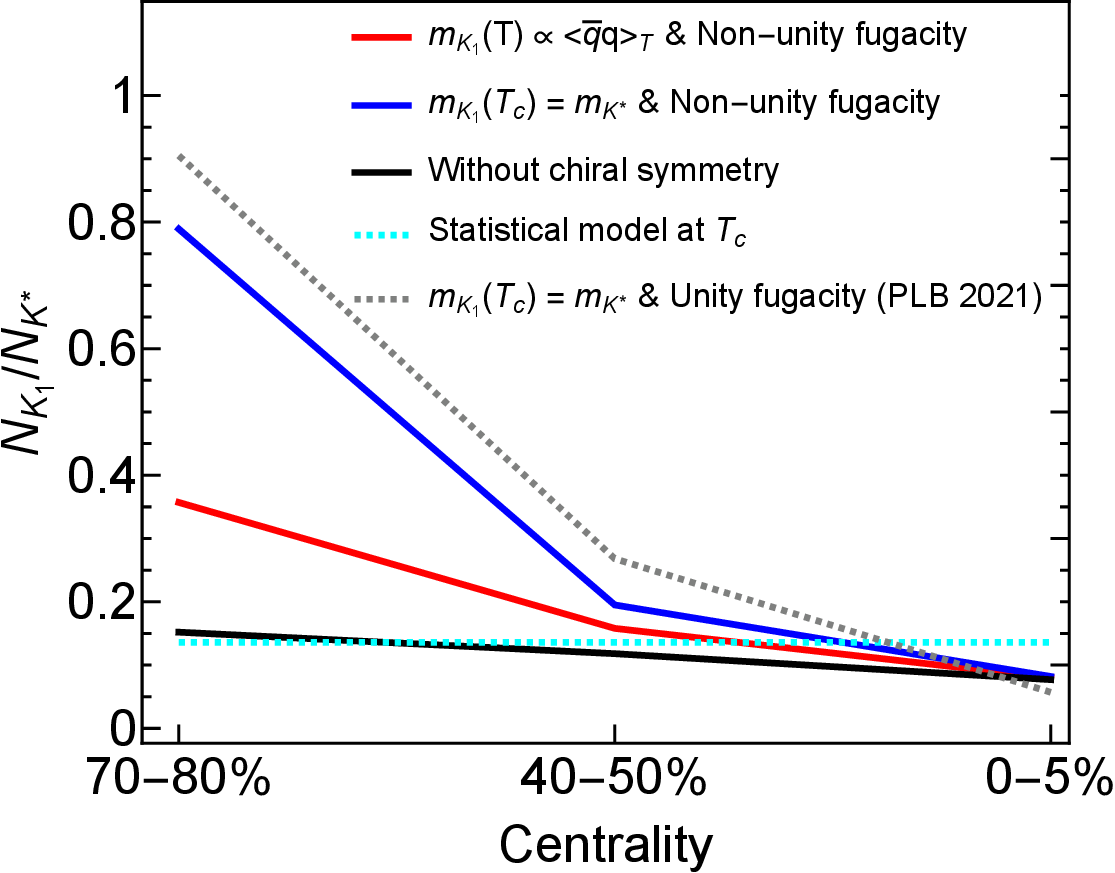}
\caption{The yield ratio $K_1/K^*$ in Pb+Pb collisions at $\sqrt{s_{NN}}=5.02$ TeV at three centralities of 0-5\%, 40-50\% and 70-80\% for various scenarios.}
\label{ratio}
\end{figure}

We solve the kinetic equations Eqs.(\ref{Kin_K1}) and (\ref{Kin_Ks}) in Sec.~\ref{kinetic} using the thermal averaged $K_1$ and $K^*$ reaction cross sections and $K_1$ decay widths given in Sec.~\ref{crosssection} and the thermal averaged $K^*$ and $K$ reaction cross sections and $K^*$ decay width from Ref.~\cite{Sung:2021myr}.  For the time dependence of the temperature of the hadronic matter after the QGP to HM phase transition in Pb+Pb collisions at $\sqrt{s_{NN}}=5.02$ TeV, we take it from Ref.~\cite{Sung:2021myr} based on a schematic ideal hydrodynamics with an equation of state from the LQCD~\cite{Borsanyi:2010cj}. Although keeping constant entropy per particle as in the present study automatically takes into account the strong viscous effect in the hadronic matter because of the increase of total particle number from the decay of resonances, it has been shown in Ref. ~\cite{Song:2010fk} that adding viscosity in the expanding hadronic matter does not affect much the time evolution of the temperature of the hadronic matter.  With an initial chemical freeze-out temperature $T_C=156$ MeV as in  Ref.~\cite{Sung:2021myr} and the initial volume of 6,076 fm$^3$, 938 fm$^3$, and 135 fm$^3$ from Ref.~\cite{Sung:2021myr} for the three collision centralities of 0-5\%, 40-50\% and 70-80\%, respectively, the effective pion, kaon, and nucleon numbers, which remain unchanged during the hadronic evolution in our study, agree with those measured by the ALICE Collaboration~\cite{ALICE:2019hno}. For the kinetic freeze-out temperatures, we take their values to be 90 MeV, 108 MeV, and 147 MeV, respectively, for the three centralities 0-5\%, 40-50\% and 70-80\% according to a blast wave model fit to the measured particle transverse momentum spectra by the ALICE Collaboration~\cite{ALICE:2019hno}. 

In Fig.~\ref{ratio}, we show the yield ratio $K_1/K^*$ from the solutions of the kinetic equations. Results including both the effect of non-unity pion and kaon fugacities as well as the temperature-dependent $K_1$ mass are shown by the solid red line with $K_1/K^*$ having values of 0.357 for peripheral collisions, 0.158 for mid-central collisions, and 0.08 for central collisions.  Compared to the results of Ref.~\cite{Sung:2021myr}, shown by the gray dashed line, in which both pion and kaon fugacities are taken to be one and the $K_1$ has a mass equal to the $K^*$ mass at $T_C$ and free-space mass below $T_C$, the final $K_1/K^*$ ratio from present study is a factor of 2.5 smaller for 70-80\% collision centrality, a factor of 1.7 smaller for 40-50\% collision centrality and a factor of 1.4 larger for 0-5\% collisions centrality. Although the collision centrality dependence of the $K_1/K^*$ ratio from the present study is thus weaker than that in Ref.~\cite{Sung:2021myr}, it still shows an enhancement in peripheral and mid-central collisions compared to the case without including the chiral symmetry restoration effect shown by the black line, indicating that an enhanced $K_1/K^*$ yield ratio in relativistic heavy ion collisions at these collision centralities remains a good signature for the chiral symmetry restoration.  We note that the reduced $K_1/K^*$ ratio in peripheral collisions in the present study compared to that in Ref.~\cite{Sung:2021myr} is mainly due to the use of more realistic temperature-dependent $K_1$ mass.  As shown by the solid blue line, without the latter effect, the non-unity pion and kaon fugacities gives a $K_1/K^*$ ratio that is only about 13\% smaller in peripheral collisions compared to the results from Ref.~\cite{Sung:2021myr}.  Also shown in Fig.~\ref{ratio} by the dashed cyan line is the $K_1/K^*$ ratio from the statistical model, which is determined at $T_{C}$ and has a value of about 0.14 independent of the collision centrality.  

We would like to point out that among the many terms in the kinetic equations for the $K_1$ and $K^*$ numbers during the  hadronic evolution, the dominant terms are those involving the $K_1$ and $K^*$ decay widths, i.e., $\langle\Gamma_{K_1\to K^*\pi}\rangle$, $\langle\Gamma_{K_1\to K \rho}\rangle$, and $\langle\Gamma_{K^*\to K\pi}\rangle$, and the thermal average of their reverse processes. Including only these terms increases the $K_1/K^*$ yield ratio by at most 18\% in essentially all considered scenarios and collision centralities. Since the width of $\rho$ meson at finite temperature is known to be significantly broadened~\cite{Rapp:1999ej}, the $K_1$ width would become larger after this effect is taken into account.
In the limit of very large $\rho$ meson width and thus large $K_1$ width, the $K_1/K^*$ ratio would approach the thermal limit given by the kinetic freeze-out temperature $T_K$.  For the scenario of non-unity fugacities and temperature-dependent $K_1$ mass considered in the present study, the $K_1/K^*$ ratio in this limit is 0.182 for peripheral collisions, 0.084 for mid-central collisions, and 0.053 for central collisions, which are all smaller than corresponding values shown in Fig.~\ref{ratio} from solving the kinetic equations as expected.  Compared to the case without the chiral symmetry restoration effect, which has the values of 0.151, 0.124, and 0.081 for peripheral, mid-central and central collisions, respectively, the $K_1/K^*$ ratio is still enhanced in peripheral collisions in this limit of fast chemical equilibration. The enhanced $K_1/K^*$ ratio is thus a robust signature of the chiral symmetry restoration effect in hot dense matter produced in peripheral relativistic heavy ion collisions. 

\section{Summary}\label{summary}

In the present study, we have extended the study of Ref.~\cite{Sung:2021myr} on the use of enhanced yield ratio $K_1/K^*$ in relativistic heavy ion collisions as a probe for chiral symmetry restoration by including non-unity pion and kaon fugacities as well as the temperature-dependent $K_1$ mass in the expanding hadronic matter.  Our results show that, although including non-unity pion and kaon fugacities only slightly reduces the $K_1/K^*$ enhancement found in Ref.~\cite{Sung:2021myr} due to chiral symmetry restoration, the inclusion of the temperature-dependent $K_1$ mass leads to a substantial reduction in the $K_1/K^*$ enhancement.  However, the final $K_1/K^*$ ratio in peripheral collisions still shows a factor of 2.4 enhancement  compared to the case without chiral symmetry restoration.  The present study thus confirms the conclusion of Ref.~\cite{Sung:2021myr} that the enhanced $K_1/K^*$ ratio can be used as a signature for chiral symmetry restoration in the hot dense matter produced in ultra-relativistic heavy-ion collisions. 

\section*{ACKNOWLEDGEMENTS}

This work was supported by the Korea National Research Foundation under Grant No. RS-2023-00280831 (S.C.), No. 2023R1A2C300302311 (S.H.L.) and Project No. NRF-2008-00458 (S.L.), and the U.S. Department of Energy under Award No. DE-SC0015266 (C.M.K.). S. H. Lee also acknowledges the support from the Samsung Science and Technology Foundation under Project No. SSTF-BA1901-04.  H. Sung thanks the Cyclotron Institute of Texas A\&M University for its hospitality during her stay as a visiting scholar supported by a graduate fellowship from the National Research Foundation of Korea under Award No. NRF-2022K1A3A1A12097807.

\bibliography{main.bib}
\end{document}